\documentclass[pre,aps,twocolumn]{revtex4}

\usepackage{graphicx}
\usepackage{amsmath}
\usepackage{amssymb}
\usepackage{ifthen}
\usepackage{booktabs}

\usepackage{graphicx}
\usepackage{dcolumn}
\usepackage{bm}
\usepackage{longtable}

\newcommand{\bb}{\begin{equation}}
\newcommand{\ee}{\end{equation}}
\newcommand{\ba}{\begin{eqnarray*}}
\newcommand{\ea}{\end{eqnarray*}}
\newcommand{\rhor}{\rho({\bf r})}
\newcommand{\dd}{{\rm d}}
\newcommand{\rr}{{\mathbf r}}
\newcommand{\dr}{{\rm d}{\bf r}}

\begin{document}

\title{Scaling behaviour of thin films on chemically heterogenous walls}

\author{Alexandr \surname{Malijevsk\'y}}
\affiliation{
{Department of Physical Chemistry, University of Chemical Technology Prague, Praha 6, 166 28, Czech Republic;}\\
 {Institute of Chemical Process Fundamentals of the Czech Academy of Sciences, v. v. i., 165 02 Prague 6, Czech Republic}}
 \author{Andrew O. \surname{Parry}}
\affiliation{Department of Mathematics, Imperial College London, London SW7 2B7, UK}
\author{Martin \surname{Posp\'\i\v sil}}
\affiliation{ {Department of Physical Chemistry, University of Chemical Technology Prague, Praha 6, 166 28, Czech Republic}}

\begin{abstract}
We study the adsorption of a fluid in the grand canonical ensemble occurring at a planar heterogeneous wall which is decorated with a chemical stripe
of width $L$. We suppose that the material of the stripe strongly preferentially adsorbs the liquid in contrast to the outer material which is only
partially wet. This competition leads to the nucleation of a droplet of liquid on the stripe, the height $h_m$ and shape of which (at bulk two-phase
coexistence) has been predicted previously using mesoscopic interfacial Hamiltonian theory. We test these predictions using a microscopic Fundamental
Measure Density Functional Theory which incorporates short-ranged fluid-fluid and fully long-ranged wall-fluid interactions. Our model functional
accurately describes packing effects not captured by the interfacial Hamiltonian but still we show that there is excellent agreement with the
predictions $h_m\approx L^{1/2}$ and for the scaled circular shape of the drop even for $L$ as small as $50$ molecular diameters. For smaller stripes
the droplet height is considerably lower than that predicted by the mesoscopic interfacial theory. Phase transitions for droplet configurations
occurring on substrates with multiple stripes are also discussed.
\end{abstract}

\maketitle

\section{Introduction}

Complete wetting is a well known example of a surface phase transition occurring at a wall-gas interface in systems for which the contact angle of a
liquid drop is zero. For open systems the transition corresponds to the growth in the thickness $\ell_{\rm eq}$ of an intruding liquid layer which
becomes macroscopic as the chemical potential $\mu$ is increased towards its saturation value; $\delta\mu=\mu_{\rm sat}-\mu\to 0$. Studies of
complete wetting date back to the Russian school of Derjaguin \cite{derj} who formulated the problem using the concept of a disjoining pressure and
were the first to recognise the importance of microscopic forces in determining the growth of the film thickness. This phase transition was
subsequently rediscovered and placed within the greater context of wetting transitions and fluid interfacial phenomena following the seminal works of
Cahn \cite{Cahn} and Ebner and Saam \cite{ES} in the 1970s. Consequently over the last three decades there have been numerous theoretical (and
experimental) studies of complete wetting at planar, and chemically homogeneous, substrates using the modern statistical mechanical theory of
inhomogeneous fluids; for reviews see for example \cite{dietrich, sullivan, schick, bonn}. Of particular importance in the theoretical development
have been treatments based on microscopic density functional theory (DFT) \cite{ev_tar} and interfacial Hamiltonian approaches which allow for the
influence of interfacial fluctuations\cite{lip}. These studies have highlighted that in addition to the growth of $\ell_{\rm eq}$, complete wetting
is also characterised by the divergence of a parallel correlation length $\xi_\parallel$ due to (thermal) fluctuations of the liquid-gas interface as
it unbinds from the wall. Renormalization group studies of interfacial models show for systems with long-ranged forces the upper critical dimension
for complete wetting $d^*<3$, meaning that mean-field treatments, including the predictions of the original Derjaguin theory, are correct in three
dimensions. More specifically, for systems with long-range attractive molecular forces decaying asymptotically as $r^{-p-4}$, the wetting film
$\ell_{\rm eq}$ and parallel correlation length $\xi_\parallel$ diverge according to the power-laws \cite{lip}
 \bb
 \ell_{\rm eq}\sim \delta\mu^{-\beta^{\rm co}},\hspace{0.5cm}\xi_\parallel\sim \delta\mu^{-\nu_\parallel^{\rm co}}
 \ee
with $\beta^{\rm co}=1/(p+1)$ and $\nu_\parallel^{\rm co}=(p+2)/2(p+1)$. Thus for systems with non-retarded van der Waals forces ($p=2$), the values
of the critical exponents are $\beta^{\rm co}=1/3$ and $\nu_\parallel^{\rm co}=2/3$, respectively.

More recently, similar theoretical tools have been used to study the adsorption and wetting on both heterogeneous (chemically patterned) and
structured solid substrates. This research has revealed a number of novel interfacial phenomena, not present for planar homogeneous substrates,
including new phase transitions and enhanced fluctuation effects. These and related studies are also of practical relevance to nanotechnologies
involving the fabrication of functional surfaces which control the adsorption of microscopically small amounts of liquid \cite{lito}. For
geometrically sculpted substrates, theoretical studies of wedges \cite{wedge_napior, wedge_parry, wedge_mal}, grooves \cite{groove_tas, groove_parry,
groove_mal} and cones \cite{cone} have shown that the wall geometry can dramatically alter the adsorption properties and accompanying interfacial
fluctuation effects. New transitions also arise for planar but chemically heterogeneous substrates in which the wall is a composite formed by
materials with different wetting properties. These include studies of unbending transitions \cite{bauer, bauer2, rascon} involving the local
condensation of liquid within a patterned region, and also complete pre-wetting (also called step-wetting) in which a nucleated liquid phase spreads
out laterally across the substrate \cite{saam,petr17}. It is even possible to combine chemical patterning and substrate geometry to produces surfaces
that are partially wet (i.e. have a finite contact angle) even though the materials involved only show complete wetting \cite{ourPRL2}. In the
canonical ensemble, i.e. considering fixed amount of liquid, chemically patterned surfaces also lead to morphological phase transition and the break
down of Young's equation \cite{lenz, gau}

In the present work we study fluid adsorption in an open system at a very simple type of chemically heterogeneous substrate; when a single stripe, of
finite width $L$ but macroscopic length and depth, of a material that prefers complete wetting is embedded into a surface that is otherwise
completely dry (or more generally, partial wet). This contrast leads to the nucleation of a droplet of liquid at the stripe the height $h_m$ of which
remains finite even at bulk two-phase coexistence ($\mu=\mu_{\rm sat}$). For this system effective Hamiltonian theories make a number of predictions
not only for the dependence of $h_m$ on $L$ but also for the precise cross-sectional shape of the drop \cite{conf}. For systems with short-ranged
forces the droplet shape is further predicted to display a conformal invariance which allows its determination on a wide variety of patterned
surfaces not just a stripe. For systems with long-ranged forces, pertinent to most solid-fluid interfaces, conformal invariance does not apply in
three-dimensions but the droplet height and shape are still predicted to show strong scaling behaviour. In the present study we test these specific
scaling predictions using an accurate density functional model based on Rosenfeld's fundamental measure theory \cite{ros}. In particular, we wish to
determine what size of stripe width $L$ is required before the droplet shape agrees with the effective Hamiltonian theory. We note that gravity is
unimportant when the stripe width $L$ is considerably smaller than the capillary length (about a mm) and is therefore utterly negligible at the
microscopic scales studied here.

The remainder of our paper is organised as follows. In section II we derive the net potential for our single-stripe wall assuming that the wall atoms
interact with the fluid via a Lennard-Jones $12$-$6$ potential. In section III we present the microscopic DFT model and also recall the interface
Hamiltonian theory for the scaling of the drop size and shape. In section IV we present our DFT results making comparison with the analytic results
of the interface Hamiltonian theory. We conclude with a discussion of our results and also highlight possible extension of this work to new phase
transitions occurring on multiple stripes.

\section{The external potential of the composite wall}

We suppose the composite wall is made from two species, $i=1,2$, each interacting with the fluid particles via the Lennard-Jones $12$-$6$ potential
 \bb
 \phi_i(r)=4\varepsilon_{i}\left[\left(\frac{\sigma}{r}\right)^{12}-\left(\frac{\sigma}{r}\right)^{6}\right]\,.
 \ee
where the molecular diameter $\sigma$ and also the uniform density distribution $\rho_w$ are considered to be identical for both species. The wall
occupies the semi-infinite volume $z<0$ and is considered macroscopic and translationally invariant along the $y$ direction. The stripe (species
$i=2$) occupies the region $|x|<\frac{L}{2}$ but the whole depth $z<0$ and length $-\infty<y<\infty$ of the wall. The total external potential
$V(x,z)$ is independent of $y$ and is obtained by integrating the wall-fluid pair potential over the wall volume from each region (see Fig.~1).  Thus
we write
 \bb
  V(x,z)=V_\infty(x,z;1)+V_L(x,z;2)+V_\infty(L-x,z;1) \label{pot}
 \ee
 The contribution due to outer parts of the wall, formed from species $i=1$, is described by the triple integral
\bb
 V_\infty(x,z;1)=\rho_w\int_x^\infty\dd x'\int_{-\infty}^\infty\dd y'\int_z^\infty \phi_1\left(\sqrt{x'^2+y'^2+z'^2}\right)\,.
 \ee
 This reduces to
  \bb
  V_\infty(x,z;1)=\pi\varepsilon_1\rho_w\sigma^3\left[\frac{\sigma^{9}}{z^9}F_9\left(\frac{x}{z}\right)
  -\frac{\sigma^{3}}{z^3}F_3\left(\frac{x}{z}\right)\right]
  \ee
  where
   \begin{widetext}
  \bb
  F_9(\xi)=\frac{2}{45}\left(1+\frac{1}{\xi^9}\right)-\frac{1}{2880}\frac {128\,{\xi}^{16}+448\,{\xi}^{14}+560\,{\xi}^{12}+280\,{
\xi}^{10}+35\,{\xi}^{8}+280\,{\xi}^{6}+560\,{\xi}^{4}+{\xi}^{2}+128}{{\xi}^{9} \left(1+{\xi}^{2}\right)^{7/2}}
  \ee
   \end{widetext}
   and
  \bb
  F_3(\xi)=\frac{1}{3}\left[1+\frac{1}{\xi^3}-{\frac {2\,{\xi}^{4}+{\xi}^{2}+2}{2{\xi}^{3} \sqrt {1+{\xi}^{2}}}}\right]\,.
  \ee

 The contribution to the potential due to the stripe, formed by species $i=2$, is given by
\begin{eqnarray}
 V_L(x,z;2)=&&\rho_w\int_{x-L}^L\dd x'\int_{-\infty}^\infty\dd y'\int_z^\infty \dd z'\nonumber\times\\
            &&\times\phi_2\left(\sqrt{x'^2+y'^2+z'^2}\right)\,.
 \end{eqnarray}
 which has now, owing to the finite size character of the stripe, the scaling form
 \bb
  V_L(x,z;2)=\pi\varepsilon_2\rho_w\sigma^3\left[\frac{\sigma^{9}}{z^9}G_9\left(\frac{x}{z},\frac{L}{z}\right)
  -\frac{\sigma^{3}}{z^3}G_3\left(\frac{x}{z},\frac{L}{z}\right)\right]
 \ee
  with
 \bb
 G_9\left(\xi,\eta\right)=F_9(\xi-\eta)-F_9(\xi)
 \ee
 and
  \bb
 G_3\left(\xi,\eta\right)=F_3(\xi-\eta)-F_3(\xi)\,.
 \ee

Note that in the limit of $L\to\infty$, the attractive part of the wall potential decays as
 \bb
V(x,z)\sim-\frac{2}{3}\pi\varepsilon_2\rho_w\sigma^3\left(\frac{\sigma}{z}\right)^3\,
 \ee
recovering the behaviour expected for a planar three dimensional homogeneous wall. On the other hand for finite $L$ and $z/L\to\infty$ the wall the
potential crosses over to
 \bb
V(x,z)\sim-\frac{3}{8}\pi\varepsilon_2\rho_w\sigma^3\left(\frac{\sigma}{z}\right)^3\frac{L}{z}\,
 \ee
which shows the correct power-law appropriate for a pseudo two-dimensional substrate.

\section{Theory}

\subsection{Density functional theory}

Within classical density functional theory \cite{evans79}, the equilibrium density profile is obtained from minimising  the grand potential
functional
 \bb
 \Omega[\rho]={\cal F}[\rho]+\int\dd\rr\rhor[V(\rr)-\mu]\,,\label{om}
 \ee
where $\mu$ is the chemical potential, and $V(\rr)$ is the external potential due to the wall. The intrinsic free energy functional ${\cal F}[\rho]$
is conveniently separated into an exact ideal gas contribution and an excess part:
  \bb
  {\cal F}[\rho]=\beta^{-1}\int\dr\rho(\rr)\left[\ln(\rhor\Lambda^3)-1\right]+{\cal F}_{\rm ex}[\rho]\,,
  \ee
where $\Lambda$ is the thermal de Broglie wavelength and $\beta=1/k_BT$ is the inverse temperature. Following the spirit of van der Waals or
equivalently simple perturbation theory, the excess part is modelled as a sum of hard-sphere and attractive contributions where the latter is treated
in a simple mean-field fashion:
  \bb
  {\cal F}_{\rm ex}[\rho]={\cal F}_{\rm hs}[\rho]+\frac{1}{2}\int\int\dd\rr\dd\rr'\rhor\rho(\rr')u_{\rm a}(|\rr-\rr'|)\,, \label{f}
  \ee
where  $u_{\rm a}(r)$ is the attractive part of the fluid-fluid interaction potential. In our model the fluid atoms are assumed to interact with one
another via the truncated (i.e., short-ranged) and non-shifted Lennard-Jones-like potential
 \bb
 u_{\rm a}(r)=\left\{\begin{array}{cc}
 0\,;&r<\sigma\,,\\
-4\varepsilon\left(\frac{\sigma}{r}\right)^6\,;& \sigma<r<r_c\,,\\
0\,;&r>r_c\,.
\end{array}\right.\label{ua}
 \ee
which is cut-off at $r_c=2.5\,\sigma$, where $\sigma$ is the hard-sphere diameter.

The hard-sphere part of the excess free energy is approximated using the Fundamental Measure Theory functional \cite{ros},
 \bb
{\cal F}_{\rm hs}[\rho]=\frac{1}{\beta}\int\dd\rr\,\Phi(\{n_\alpha\})\,,\label{fmt}
 \ee
which accurately takes into account short-range correlations between fluid particles and is known to satisfy exact statistical mechanical sum rules
\cite{hend}. Here, the free energy density $\Phi$ depends on a set of six weighted densities $\{n_\alpha(\rr)\}$ which may be expressed as double
integrals over the $x$ and $z$ dimensions \cite{fmt_mal}. These are evaluated numerically using Gaussian quadrature.

Minimisation of (\ref{om}) leads to the Euler-Lagrange equation
 \bb
 V(\rr)+\frac{\delta{\cal F}_{\rm hs}[\rho]}{\delta\rho(\rr)}+\int\dd\rr'\rho(\rr')u_{\rm a}(|\rr-\rr'|)=\mu\,,\label{el}
 \ee
which can be solved iteratively on an appropriately discretized two dimensional grid with suitable boundary conditions (see later).

\subsection{Interfacial Hamiltonian theory}

An alternative and complementary approach to microscopic DFT, applicable to interfacial phenomena occurring at the mesoscopic scale is based on the
analysis of simple effective Hamiltonian models. These have been  successfully applied to the theory of wetting at planar (homogeneous) walls
\cite{forgacs, lip} and also continuous wedge filling \cite{wedge_parry} providing a complete classification of universality classes and scaling
regimes arising due to the interplay between fluctuation effects and those directly coming from the intermolecular forces.  In application to the
present heterogeneous wall, the very simplest effective Hamiltonian that can be used is given by the model \cite{conf}
 \bb
 H[h]=\int\dd x\left\{\frac{\gamma}{2}\left(\frac{d h}{d x}\right)^2+W(h)\right\}\,,\label{Hh}
 \ee
describing the functional dependence of the free-energy (per unit wall length) on the height $h(x)$ of the liquid-vapour interface $h(x)$ above the
stripe. Here $\gamma$ is the liquid-vapour surface tension and $W(h)$ is the binding potential describing the complete wetting properties which we
specify below. At mean field level the equilibrium configuration of the interface is be determined by simply minimising the functional (\ref{Hh}),
yielding the Euler-Lagrange equation
  \bb
  \gamma\frac{d^2 h}{d x^2}=W'(h)\,. \label{EL}
  \ee
This must be solved subject to boundary condition $h(|L|/2)=h_e$ where the edge value $h_e$ is considered fixed and microscopic.

\begin{figure}[h]
\centerline{\includegraphics[width=8cm]{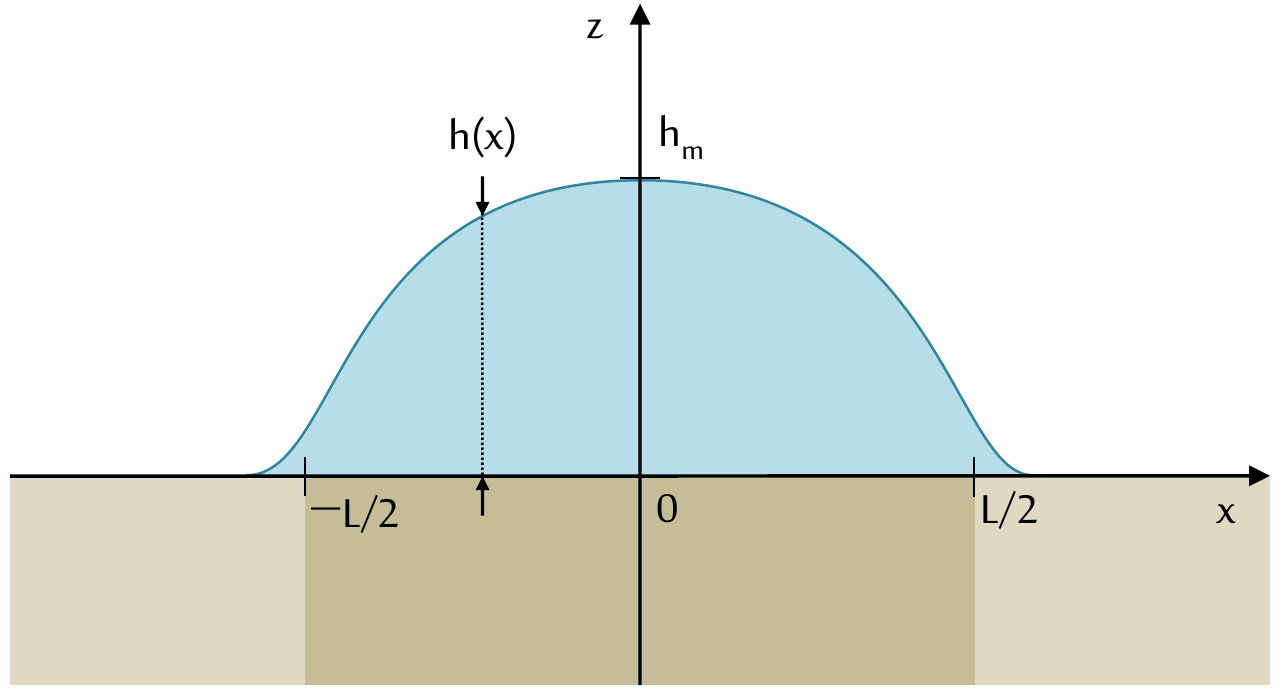}}
 \caption{Schematic illustration of the shape of a drop, of height $h_m$, above a composite planar
wall containing stripe of material, of finite width $L$ but infinite depth and length, which is completely wet. The outer parts of the substrate only
show microscopic adsorption corresponding to partial wetting (or indeed drying).}
\end{figure}

Within the model (\ref{Hh}) the width of the stripe $L$ is assumed to be large enough that the binding potential can be considered to be independent
of $x$ and approximated by the same function appropriate to wetting at planar homogeneous wall. In the case of the Lennard-Jones wall-fluid
interaction the binding potential, at bulk saturation, takes the form
 \bb
W(h)=\frac{A}{h^2}\,,\label{bind}
 \ee
where $A=\pi\varepsilon_2\rho_w\sigma^6(\rho_l-\rho_v)/3$ is the Hamaker constant associated with the wettable part of the wall and $\rho_l$ and
$\rho_v$ are the liquid and vapour saturated densities, respectively. Substituting (\ref{bind}) into (\ref{EL}) and integrating once we obtain (for
$x<0$)
 \bb
 \frac{d h(x)}{d x}=\sqrt{\frac{2A}{\gamma}}\frac{\sqrt{h_m^2-h^2(x)}}{h_mh(x)} \label{hx}
 \ee
and with a change of sign for $x>0$.  Here $h_m$ is the maximum of $h(x)$ occurring at $x=0$. Further integration leads to
 \bb
 \sqrt{h_m^2-h^2(x)}=\sqrt{\frac{2A}{\gamma}}\frac{x}{h_m}\,, \label{el2}
 \ee
and hence from the boundary condition
 \bb
 \sqrt{h_m^2-h_e^2}=\sqrt{\frac{2A}{\gamma}}\frac{L}{2h_m}\,.
 \ee
Assuming that $h_m\gg h_e$ we obtain the simple power-law dependence of the droplet height on the stripe width
  \bb
  h_m^2\approx L\sqrt{\frac{2A}{\gamma}}\,. \label{hm}
  \ee
which is independent of $h_e$ and hence on the properties of the outer wall (provided these remain partial wet). Substituting (\ref{hm}) into
(\ref{el2}) then implies that the droplet height has a simple universal circular shape
   \bb
  \tilde{h}(x)=\sqrt{1-4\tilde{x}^2}\,, \label{scale}
   \ee
  when expressed in re-scaled variables $\tilde{x}=x/L$ and $\tilde{h}(x)=h(x)/h_m$.

The square root power-law dependence of $h_m$ on $L$ is in accord with expectations based on simple scaling theory. In the presence of more general
intermolecular forces finite scaling suggests that for very wide stripes and away from co-existence the droplet height scales as $h_m\approx
\delta\mu^{-\beta^{\rm co}} H(L/\xi_\parallel)$ where recall $\xi_\parallel\approx\delta\mu^{-\nu_\parallel^{\rm co}}$ is the complete wetting
parallel correlation length. For fixed $L$ and $\delta\mu\to 0$ we require the scaling function vanishes as $H(x)\propto x^{\beta^{\rm
co}/\nu_\parallel^{\rm co}}$ implying $h_m\propto L^\frac{2}{p+2}$ recovering (\ref{hm}) for the present case of van der Waals forces ($p=2$).

For this model it is also possible to determine the equilibrium free-energy (per unit stripe length) of the droplet. Substituting the equilibrium
profile back into (\ref{Hh}) determines that
  \bb
  F=(\gamma + W(h_m))L+\sqrt{8\gamma}\int_{h_e}^{h_m} dh \sqrt{W(h)-W(h_m)}
   \ee
where we have also included the extensive contribution $\gamma L$ coming from the interfacial tension. For the binding potential (\ref{bind}) the
integral can be evaluated quite easily giving
 \begin{eqnarray}
  F&=&(\gamma + W(h_m))L\\
  &&-\sqrt{8\gamma
  A}\left[\sqrt{1-\frac{h_e^2}{h_m^2}}+\ln\left(\frac{h_m}{h_e}-\sqrt{\frac{h_m^2}{h_e^2}-1}\right)\right]\nonumber
   \end{eqnarray}
 For large $L/h_e\gg 1$ this has the expansion
 \bb
  F=\gamma L+\sqrt{2 A\gamma}\ln\frac{L}{h_e}+\cdots
  \label{Fsing}
   \ee
where the higher-order terms are of order unity and also depend on $h_e$,$\gamma$ and $A$. The presence of the logarithmically diverging Casimir term
means it is not possible to define a thermodynamic line tension associated with the three phase contact line for this system. Again this term is
consistent with the anticipated finite-size scaling of the free-energy at complete wetting. Recall that for complete wetting at an infinite uniform
planar wall, the wall-gas interfacial tension contains a singular contribution $\gamma_{\rm sing}\approx \delta\mu^{2-\alpha_s^{co}}$ where standard
exponent relations determine that $2-\alpha_s^{co}=1-\beta^{\rm co}$ \cite{schick}. For a droplet of width $L$ we expect that this contribution
scales as $\gamma_{\rm sing}\approx \delta\mu^{2-\alpha_s^{co}}B(L/\xi_\parallel)$ with $B(x)$ a suitable scaling function. In the limit of
$\delta\mu\to 0$ this implies that the free-energy per unit length (along the $y$-axis) of the drop should contain a singular contribution $F_{\rm
sing}\propto L^{1-(2-\alpha_s^{co})/\nu_\parallel^{\rm co}}$ which reduces to $F_{\rm sing}\propto L^{(2-p)/(2+p)}$. This diverges as $L\to \infty$
for $p<2$ which is consistent with the marginal logarithmic divergence shown in (\ref{Fsing}) for the present case $p=2$. Similar logarithmic
contributions occur in two dimensions for systems with short-ranged forces where they arise due to fluctuation effects. This similarity is not
coincidental since in this dimension interfacial wandering leads to an effective entropic repulsion which also decays as an inverse square
\cite{jak,abraham}.

\begin{figure}[h]
\centerline{\includegraphics[width=8cm]{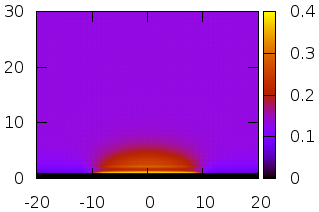}}
 \centerline{\includegraphics[width=8cm]{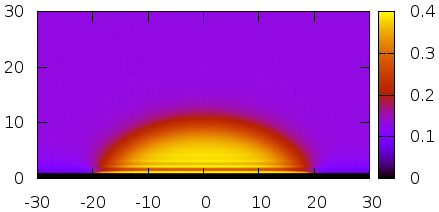}}
 \centerline{\includegraphics[width=8cm]{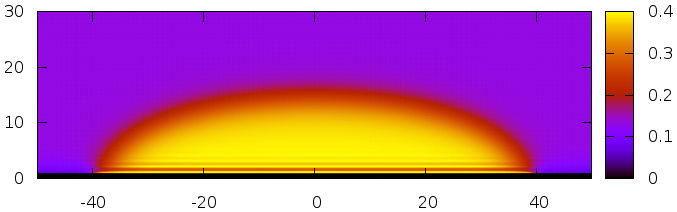}}
 \caption{Equilibrium density profiles as obtained from DFT at a temperature $T=0.95\,T_c$ and bulk two-phase equilibrium
          for stripe widths (from top to bottom): $L=20\,\sigma$, $L=40\,\sigma$, and $L=80\,\sigma$.}\label{p_dens}
\end{figure}

\section{Results}

\begin{figure}[h]
\centerline{\includegraphics[width=10cm]{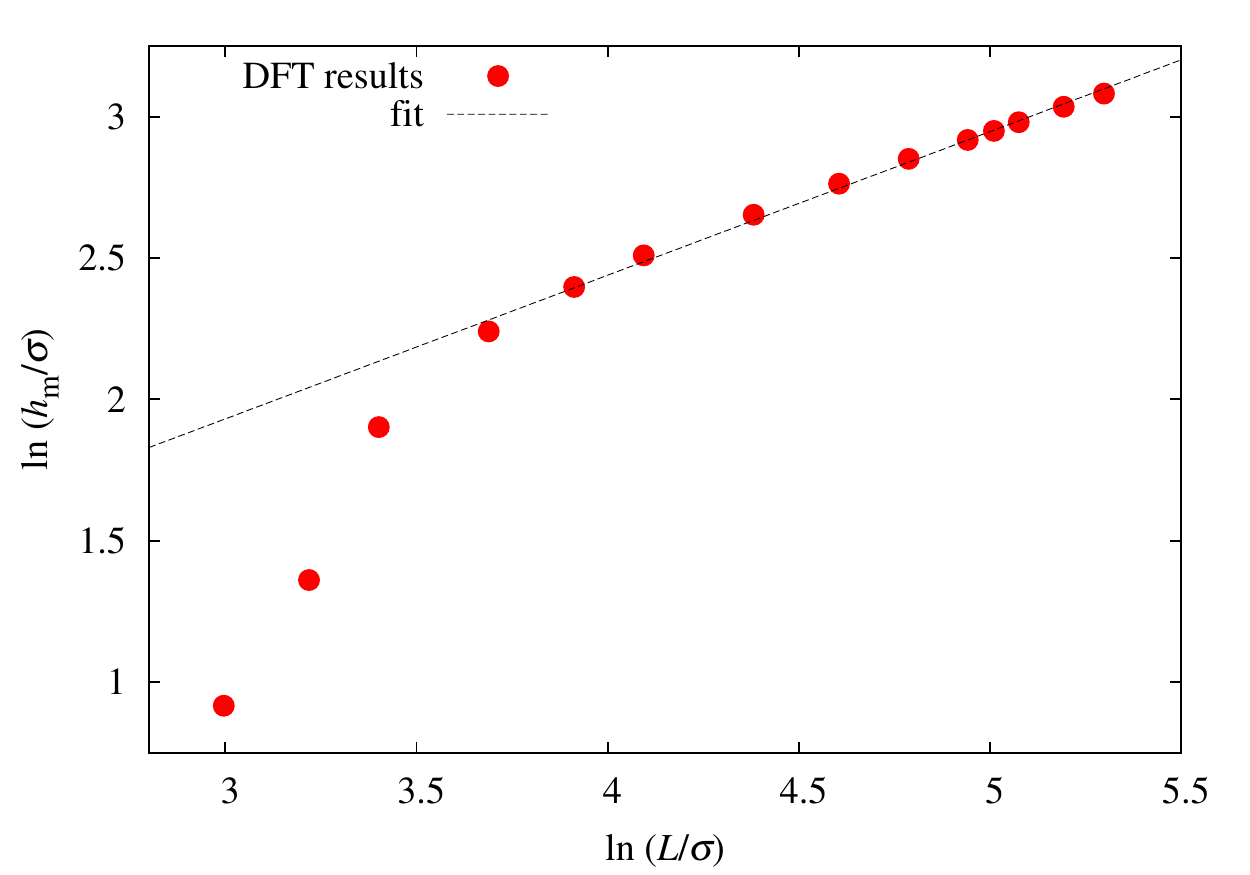}}
 \caption{Dependence of the drop height $h_m$ on the stripe width $L$. Symbols denote the DFT results with $h_m$ determined
from the equilibrium density profile using the criterion $\rho(0,h_m)=(\rho_v+\rho_l)/2)$. The straight line fit to the data for $L\ge40\,\sigma$ has
slope $0.5079$.}\label{loglog}
\end{figure}

In our DFT calculations we have considered the simplest case, setting $\varepsilon_1=0$, so that the outer material is modelled as a purely hard
wall. This part of the substrate is therefore completely dry, with local contact angle $\theta_1=\pi$, at all (sub-critical) temperatures. For the
material of the stripe we set $\varepsilon_2\rho_w=1.2\,\varepsilon\sigma^{-3}$ which, from earlier studies \cite{our_prl}, we know leads to a
first-order wetting transition occurring at $T_w/T_c=0.83$ were it of infinite extent. Here $T_c$ is the bulk critical temperature which in our model
occurs at $k_BT_c/\varepsilon=1.41$. Our present calculations are performed at $T/T_c=0.95$ far above the wetting transition ensuring that the local
contact angle of the stripe phase $\theta_2=0$. This is also sufficiently below the critical temperature that the bulk correlation length is small,
of order $\sigma$ \cite{our_prl}. Furthermore, we fix the chemical potential to its saturation value at this temperature, $\mu\equiv\mu_{\rm
sat}=-3.945\,\varepsilon$. For reference the coexisting bulk vapour and gas densities are $\rho_v=0.1308\sigma^{-3}$ and $\rho_l=0.3872\sigma^{-3}$,
respectively. The Euler-Lagrange equation (\ref{el}) was solved numerically on a 2D rectangular grid with mesh spacing $0.1\,\sigma$ (an order of
magnitude smaller than the bulk correlation length) using standard Picard iteration. We considered a box of overall width $L_x=250\,\sigma$ and
height $L_z=30\,\sigma$ and determined the equilibrium density profiles for different stripe widths $L=10\,\sigma,20\,\sigma\cdots,200\,\sigma$. As
can be seen from the illustrative density profiles displayed in Fig.~\ref{p_dens}, a drop-like structure forms above the stripe for $L>20\,\sigma$.
For smaller widths only a microscopic film forms above the stripe and the mesoscopic interfacial Hamiltonian theory is not applicable. For larger
widths however it is possible to extract the shape of an equilibrium drop $h(x)$ from the density profile $\rho(x,z)$ by tracing the contour of
iso-density $\rho(x,h(x))=(\rho_v+\rho_l)/2$. From this we can, in particular, extract the maximum (mid-point) height
$\rho(0,h_m)=(\rho_v+\rho_l)/2$. In Fig.~\ref{loglog} we show a log-log plot illustrating the dependence of $h_m$ on the stripe width $L$. For
$L>40\,\sigma$ there is evidently a linear dependence for which we find the slope  $0.5079\pm0.01$ which is in excellent agreement with the
prediction of Eq.~(\ref{hm}) of the effective Hamiltonian theory. Interestingly, there is a sharp cross-over in the form of $h_m(L)$ at smaller
values of $L$ which we discuss later.

\begin{figure}[h]
\centerline{\includegraphics[width=10cm]{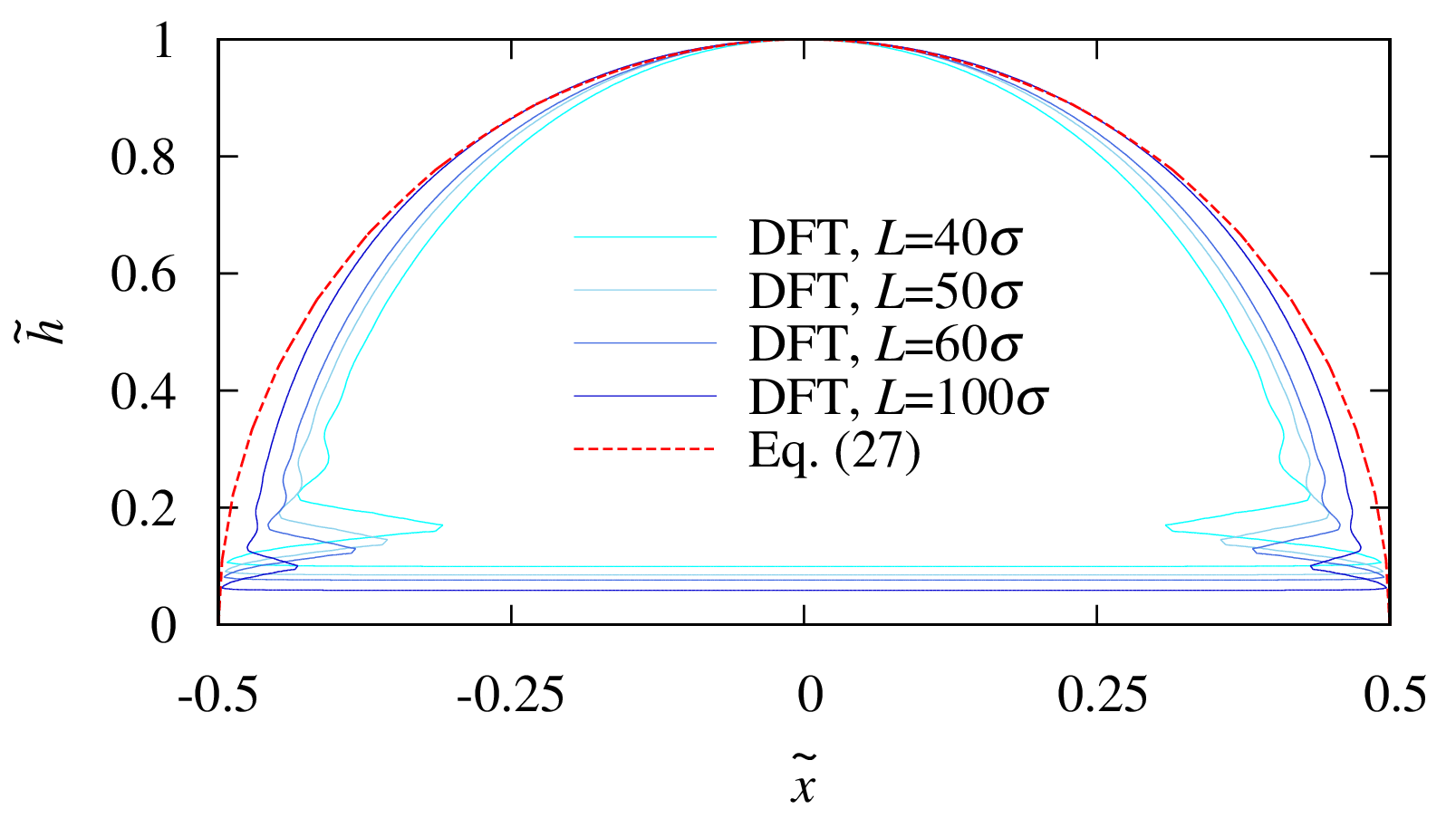}}
 \caption{DFT results for the droplet shape for different stripe widths $L$ in rescaled dimensionless units $\tilde{x}=x/L$ and $\tilde{h}(x)=h(x)/h_m$. The dashed red line is the prediction of the effective
 Hamiltonian model.}\label{p_scale}
\end{figure}

Finally, in Fig.~\ref{p_scale} we show our numerical results for the drop shape $h(x)$ in suitable dimensionless rescaled units $\tilde{x}=x/L$ and
$\tilde{h}(x)=h(x)/h_m$ for different system sizes. While there is not perfect data collapse there is clear indication that even for relatively small
system sizes the rescaled shape of the droplet is close to, and converging towards, the analytic result given by Eq.~(\ref{scale}) (dashed red line).

\section{Summary}

In this work we have used an accurate DFT based on a Rosenfeld-like fundamental measure theory model to study the formation and shape of liquid drops
on a composite planar wall decorated by a single stripe of lateral width $L$ which is completely wet. For both the mid-point height and droplet shape
we have found very good agreement with the predictions of effective Hamiltonian theory. Indeed we were surprised that the agreement is good even down
to stripes which are only $50\sigma$ across. This contrasts with the findings of DFT studies of capillary condensation where the agreement with the
macroscopic Kelvin equation, for conditions where the walls which are completely wet, only occurs if the slit width is many hundreds of molecular
diameters. This is indicative that mesoscopic effective Hamiltonians are, in general, reliable when the confining length scales are much larger than
the underlying bulk correlation length (which is molecularly small in the present study). We were also intrigued to note that there appears to be a
sharp cross-over in the dependence of $h_m$ on $L$ for $L<50\sigma$. It may well be possible to explain this behaviour by extending the simple
interfacial model to allow for a position dependent binding potential $W(h;x,L)$. Indeed within a sharp kink approximation, where we model the
density profile as a sharp step function between the bulk liquid and gas, we can determine this from the external potential using
$W(h;x,L)=-(\rho_l-\rho_v)\int_h^\infty V(x,z) dz$. We can also improve on (\ref{Hh}) by using the full expression for the interfacial area rather
than a square gradient approximation. However, given that the resulting equations must be solved numerically they seem to offer little advantage over
the full microscopic DFT calculation. Another point to note is that both approaches employed in this work were based on a mean-field approximation
which neglects the effect of interfacial fluctuations. While these should not affect the scaling results for the size and shape of larger drops it is
possible that they are important for smaller system sizes. Simulation studies of this would be very welcome.

To finish we point out that new phenomena are likely to emerge when we consider walls decorated with multiple wetting stripes on a dry (or partially
wet) surface. Consider for example the case of two parallel stripes of width $L$ separated by a distance $D$. When $D\gg L$ the stripes are
essentially independent and isolated drops, similar to those described here, form above them. As the distance $D$ is decreased the two droplets must
eventually coalesce to form a single drop containing a bubble of gas which spans the dry part of the substrate between the two stripes. This would be
an example of a bridging transition akin to those studied for fluids between spheres and cylinders \cite{hopkins, bridge}. The distance $D_B$ at
which this bridging transition occurs will be determined largely by the logarithmic Casimir interaction shown in Eq.~\ref{Fsing} \cite{fut} and we
anticipate $D_B\propto\ln L$. With three stripes bridging must also occur as we reduce the distance(s) between them but now we also have the
possibility that the coalesced drop covers two or indeed three stripes. Similarly, it is natural to imagine that on a periodic array of stripes
similar bridging transitions are induced by varying the inter-stripe gap. In this case such transitions must be accompanied by symmetry breaking
since if we label the (wet) stripes $i=1,2,3,4\cdots$ coalesced drops may span stripes $1$ to $2$, $3$ to $4$ etc or (for exactly the same
free-energy) $2$ to $3$, $4$ to $5$ etc. Indeed, it is possible that a sequence of symmetry breaking first-order transitions, producing larger
coalesced droplets, is encountered as the distance between the stripes is reduced further. This may lead to very complex phase behaviour even before
the interplay with other surface phase transitions is allowed for. We hope that the present study is a first step to classifying these potentially
very rich phase diagrams.

\begin{acknowledgments}
\noindent This work was funded in part by the EPSRC UK grant EP/L020564/1, ``Multiscale Analysis of Complex Interfacial Phenomena''.
 A.M. acknowledges the support from the Czech Science Foundation, project 17-25100S.
\end{acknowledgments}

\end{document}